\documentstyle[12pt]{article}
\begin{document}

\begin{titlepage}
\vspace{0.2in}
\begin{flushright}
DOE/ER/01545-589\\
OHSTPY-HEP-T-93-003\\
CPT-93/P.2908\\
June 1993
\end{flushright}
\vspace*{0.5cm}
\begin{center}
\begin{Large}
{\bf Analyticity Properties and Unitarity\\ Constraints
of Heavy Meson Form Factors\\}
\end{Large}
\vspace*{1.0cm}
%\vspace*{0.7cm}
Eduardo de Rafael \\
\vspace*{0.7cm}
%\vspace*{0.5cm}
Centre de Physique Th\'eorique, C.N.R.S. - Luminy,
Case 907. \\
F-13288 Marseille Cedex 9, France. \\
\vspace{1.0cm}
Josep Taron \\
\vspace*{0.7cm}
%\vspace*{0.5cm}
The Ohio State University, Department of Physics. \\
174 West 18th Avenue,\\
Columbus, OH-43210-1106, USA.\\
%\vspace*{1cm}
\vspace*{2cm}
{\bf   Abstract  \\ }
\end{center}
We derive new bounds on the b-number form factor $F(q^2)$
of the B meson
for $q^2$-values relevant to the kinematics of the decays
$\bar{B} \to D l {\bar{\nu}}_l$ and $\bar{B} \to D^* l {\bar{\nu}}_l$.
The new bounds take into account the experimentally known properties
of the $\Upsilon$-states below the onset of the
physical $\bar{B} B$-threshold.

\vfill
\end{titlepage}
\noindent
{\bf 1. INTRODUCTION.}
\vspace{0.1in}

The possibility of obtaining model independent bounds \cite{drt}
on the b-number form factor $F(q^2)$ of the B meson:
\begin{equation}
\label{i1}
<B(p')|\bar{b} \gamma^\mu b |B(p)>=(p+p')^\mu \; F[q^2=(p-p')^2],
\end{equation}
has recently attracted some attention (refs. \cite{flw} to \cite{bh}). The
interest of these bounds for phenomenology lies in their relevance to the
semileptonic B-decays
\begin{equation}
\label{i2}
\bar{B} \to D l {\bar{\nu}}_l \;\;\; {\rm and} \;\;\;
\bar{B} \to D^* l {\bar{\nu}}_l.
\end{equation}
It has been shown \cite{iw} that in the limit of very large b and c quark
masses, there are new approximate symmeries of QCD which allow one to
express the six form factors which govern these decays in terms of the
b-number form-factor $F(q^2)$ in (1)
alone. The conservation of b-number by the
strong interactions implies
\begin{equation}
\label{i3}
F(q^2=0)=1.
\end{equation}
Further model independent information on this form factor; e.g. about its
slope at the origin, would be very useful to extract the value of the
mixing matrix element $|V_{cb}|$ from the data \cite{argus,bs}.

The bounds proposed in ref. \cite{drt} are based on very general QCD
properties of the two point function $\Pi (q^2)$ defined as
$$
(q^\mu q^\nu - q^2 g^{\mu \nu}) \Pi (q^2)= i \int d^4 x \; e^{iq.x} \;
<0| T \left( V^\mu (x) V^\nu (0) \right) |0>,
$$
\begin{flushright}
(4a)
\end{flushright}
with
%\newpage
$$V^\mu= \bar{b}(x) \gamma^\mu b(x) \; ;$$
\begin{flushright}
(4b)
\end{flushright}
as well as on analyticity properties which the b-number form factor
$F(q^2)$ in eq.(1) is {\it assumed} to satisfy in the large b-quark mass
limit. In QCD, the function $\Pi (q^2)$ obeys a once subtracted dispersion
relation. It is therefore convenient to consider the first derivative
of $\Pi (q^2)$
($Q^2=-q^2$, our metric is $+ - - -$, $Q^2>0$ corresponds to the
spacelike region):
\setcounter{equation}{4}
\begin{equation}
\label{i5}
\chi(Q^2)=-{{\partial \Pi (Q^2)} \over {\partial Q^2}}=
\int_0^{\infty} dt {1 \over {(t+Q^2)^2}}{1 \over \pi} Im \Pi (t),
\end{equation}
with $Im \Pi (t)$ the b-number spectral function defined by the relation
$$(q^\mu q^\nu-g^{\mu \nu} q^2) Im \Pi (q^2)=$$
\begin{equation}
\label{i6}
{1 \over 2} \sum_{\Gamma} \int d \mu (\Gamma) {(2 \pi)}^4
{\delta}^{(4)} (q-\sum_{\Gamma} p) \;
<0| V^\mu (0) |\Gamma><\Gamma| V^\nu (0)|0>,
\end{equation}
where the summation is extended to all possible hadronic states $\Gamma$
with the quantum numbers of the $V^\mu$ current, and with the integral over
phase space extended to each intermediate state.

Two bounds were derived in ref. \cite{drt}. The first bound follows from
the saturation of the r.h.s. in eq.(6) with the lowest $\bar{B} B$-state.
Since each hadronic state contributes positively to the spectral function,
we have in this case
\begin{equation}
\label{i7}
{1 \over \pi} Im \Pi (t) \geq { 1 \over {16 \pi^2}} {1 \over 3}
{\left( 1- {{4M_B^2} \over t} \right)}^{3/2} {|F(t)|}^2
\theta (t- 4M^2_B),
\end{equation}
with $F(t)$ the same b-number form factor as in (1). The second bound
takes also into account the other two-meson states: $\bar{B} B^*$,
${\bar{B}}^* B$, and ${\bar{B}}^* B^*$; {\it assuming further} that, in the
large b-quark mass limit, the four states $\bar{B} B$, $\bar{B} B^*$,
${\bar{B}}^* B$, and ${\bar{B}}^* B^*$ are related to each other by the
resulting new spin-flavour symmetry. (The assumption here is in fact
similar to the one previously made in refs. \cite{mr} and \cite{fg} to
predict the ratios of $e^+ e^- \to \bar{B} B,\bar{B} B^*,
{\bar{B}}^* B,{\bar{B}}^* B^*$ cross sections.) Needless to say, the second
bound is stronger than the first; and, when compared to the existing
model dependent determination of the $F(q^2)$ form factor
\cite{models} and the
fits to the present experimental data,
turns out to be surprisingly
restrictive \cite{jt}. This has prompted several authors to
reconsider critically some of the assumptions which were made in \cite{drt}.

The basic criticism of refs. \cite{flw,gm,dkp,cmimr}
focuses on the analyticity properties which in ref. \cite{drt}
are implicitly attributed to the b-number form factor $F(q^2)$ in the large
b-quark mass limit; and in particular the neglect of the effect of
heavy-heavy "onium" states below the $4M_B^2$-threshold. None of these
references, however, offers the derivation of new useful bounds compatible
with the {\it rectified} analyticity properties. The main purpose of this
article is to show how to derive new bounds on $F(q^2)$, for $q^2$-values
relevant to the kinematics of the decays in (2), with inclusion of the
experimentally known properties of the $\Upsilon$-states below the onset of
the physical $\bar{B} B$ threshold.

The paper is organized as follows. Section 2 reviews general positivity
properties of the b-number spectral function, as well as the analyticity
properties of the b-number form factor of the B-meson. The new bounds
on this form factor, in the presence of the $\Upsilon$-states below
the onset of the physical $\bar{B} B$-threshold, are derived in section 3.
Section 4 discusses a number of observations relevant to the Heavy Quark
Effective Theory (HQET) which stem from this work. The technical details
to derive the new bounds are explained in an appendix.
\newpage

\noindent
{\bf 2. UNITARITY CONSTRAINTS, ANALYTICITY PROPERTIES, and QCD.}
\vspace{0.1in}

The starting point is the dispersion relation in eq.(5) and the
positivity property of the b-number spectral function defined in
eq.(6). The contribution to this spectral function from each of the
$\bar{B} B$ intermediate states: $B^- B^+$, ${\bar{B}}^0 B^0$, and
${\bar{B}}^0_s B^0_s$ is the same in the limit where the light quark mass
differences are neglected. As pointed out to us by D.J.Broadhurst
(and emphasized in refs. \cite{db} and \cite{agg}), this brings a
factor $n_f$ which counts the number of light flavours in the r.h.s. of
eq.(7); i.e.,
\begin{equation}
\label{i8}
{1 \over \pi} Im \Pi (t) \geq { {n_f} \over {16 \pi^2}} {1 \over 3}
{\left( 1- {{4M_B^2} \over t} \right)}^{3/2} {|F(t)|}^2
\theta (t- 4M^2_B).
\end{equation}

If necessary,
one can improve this inequality by explicitly taking into account
the contribution to the b-number spectral function from the
$\Upsilon$-states below the two-meson threshold. (We refrain from including
the contribution of the $\Upsilon$-states above threshold as well, because
of the danger of possible double counting with the $B \bar{B}$ continuum.)
These contributions can be extracted from the $e^+ e^- \to \Upsilon$
experimental cross-section, due to the fact that the hadronic
electromagnetic current brings in the b-number current via the term
$-1/3 \; \bar{b} \gamma^\mu b$. Using the simple parametrization
\begin{equation}
\label{i9}
\sigma (e^+ e^- \to \Upsilon)= 12 \pi^2 \delta (t-M_{\Upsilon}^2)
{{\Gamma^{ee}_{\Upsilon}} \over M_{\Upsilon}},
\end{equation}
and the relation ($-e/3$ is the b-quark electric charge)
\begin{equation}
\label{i10}
\sigma (e^+ e^- \to \Upsilon)={{4\pi^2 \alpha} \over t} {{e^2} \over 9}
{1 \over \pi} Im \Pi_{\Upsilon} (t),
\end{equation}
we can improve eq.(8) to
$${1 \over \pi} Im \Pi (t) \geq  {{27} \over {4 \pi \alpha^2}} \sum_{i}
M_{\Upsilon_i} \Gamma^{ee}_{\Upsilon_i} \delta (t - M^2_{\Upsilon_i})$$
\begin{equation}
\label{i11}
+ { {n_f} \over {16 \pi^2}} {1 \over 3}
{\left( 1- {{4M_B^2} \over t} \right)}^{3/2} {|F(t)|}^2
\theta (t- 4M^2_B).
\end{equation}

Inclusion of the contribution from other two meson intermediate states to
the b-number spectral function necessarily requires additional
dynamical assumptions
at the present stage. As pointed out in ref. \cite{bds},
the HQET cannot be reliably applied
to relate the various physical amplitudes: $<\bar{B} B |V^\mu|0>$,
$<{\bar{B}}^* B |V^\mu|0>$, and
$<{\bar{B}}^* B^* |V^\mu|0>$ in the time
like region. We shall therefore limit ourselves to the derivation of a new
bound based on the inequality (11) only.

The function $\chi (Q^2)$, for a heavy quark mass $m_b$ and space like
values $Q^2 \geq 0$, can be reliably computed using QCD perturbation theory
. At the one loop level, asymptotic freedom results in ($N_c=$ number
of colours)
\begin{equation}
\label{i12}
\chi(Q^2)={{N_c} \over {4 \pi^2}} \int_0^1 dx
{{2 x^2 (1-x)^2} \over {m_b^2 + x(1-x) Q^2}} \;\;.
\end{equation}
The knowledge of this function, and the lower bound for the spectral
function in (11) inserted into the dispersion relation in (5) lead to the
unitary inequality
$$16 \pi^2 M_B^2 \chi(Q^2) \geq
{{27 \pi} \over {4 \alpha^2}}
\sum_i{{ M_{\Upsilon_i}\Gamma^{ee}_{\Upsilon_i}} \over {M_B^2}}
{\left( {{M_{\Upsilon_i}^2+Q^2} \over {4M_B^2}} \right)}^{-2}$$
\begin{equation}
\label{i13}
+
{{n_f} \over 12} \int_1^{\infty} dy
\left(y+ {{Q^2} \over {4M_B^2}}\right)^{-2}
y^{-3/2} (y-1)^{3/2} |F(4M_B^2 y)|^2,
\end{equation}
where we have set $y=t/4M_B^2$.

We next turn our attention to the b-number form factor $F(t)$ of the
B-meson. On general quantum field theory grounds $F(t)$ obeys a dispersion
relation; and it follows from the QCD inequality above that the dispersion
relation for $F(t)$ has at most one subtraction. Since the value of
$F(t)$ at t=0 is known (see eq.(3)), it is convenient to use t=0 as the
subtraction point i.e.,
\begin{equation}
\label{i14}
F(t)=F(0)+ {t \over \pi} \int_0^{\infty} {{dt'} \over {t'}}
{{ImF(t')} \over {t'-t-i\epsilon}}\;\;.
\end{equation}
In full generality
$$(p+p')^\mu ImF(t)=$$
\begin{equation}
\label{i15}
{1 \over 2} \sum_{\Gamma} \int d \mu (\Gamma) {(2 \pi)}^4
{\delta}^{(4)} (q-\sum_{\Gamma} p) \;
<\bar{B} B|\Gamma><\Gamma| V^\mu (0)|0>,
\end{equation}
with the summation extended to all possible hadronic states $\Gamma$ with
the quantum numbers of the $V^\mu$ current. It appears then that the
b-number form factor of the B-meson
has a succession of branch cuts starting at the $\pi \pi$-
threshold, the $K \bar{K}$-threshold, the $D \bar{D}$-threshold, the
$B \bar{B}$-threshold, etc. Since the $V^\mu$ current only involves b-quarks
and b-quarks are heavy, their coupling to hadronic states of lighter
flavours
-which can only proceed through annihilation via gluonic interactions- are
suppressed. Other possible hadronic states below the $B \bar{B}$-threshold
are the three $\Upsilon$-states:
$\Upsilon(1S)$, $\Upsilon(2S)$ and $\Upsilon(3S)$ which to a good
approximation appear as poles in the positive real axis of the complex
t-plane. Their contribution to the b-number form factor can be parametrized
as follows
\begin{equation}
\label{i16}
F(t)=F(0)+t \sum_i 3 {{g_{\Upsilon_i B \bar{B}}f_{\Upsilon_i}} \over
{M_{\Upsilon_i}^2-t-i \epsilon}} + F_{reg}(t),
\end{equation}
where $f_{\Upsilon_i}$ denote the coupling constants which govern the
electronic width of the $\Upsilon_i$-resonances
\begin{equation}
\label{i17}
\Gamma( \Upsilon_i \to (\gamma) \to e^+ e^-)= f^2_{\Upsilon_i}
M_{\Upsilon_i} {{4 \pi} \over 3} \alpha^2,
\end{equation}
and $g_{\Upsilon_i B \bar{B}}$ the coupling constants of the
$\Upsilon_i$-resonances to the $B \bar{B}$ system. More precisely, we are
considering an effective Lagrangian interaction
\begin{equation}
\label{i18}
{\cal L}_{\Upsilon \gamma}= {{|e|} \over 2} f_{\Upsilon} ( \partial_\mu
\Upsilon_\nu - \partial_\nu \Upsilon_\mu) \; F^{\mu \nu},
\end{equation}
to implement the coupling of a massive spin 1 field which describes a
generic $\Upsilon$-resonance, with the electromagnetic strength tensor
$F^{\mu \nu}=\partial^\mu A^\nu- \partial^\nu A^\mu$; and an effective
interaction lagrangian
\begin{equation}
\label{i19}
{\cal L}_{\Upsilon B \bar{B}}= i g_{\Upsilon B \bar{B}} \Upsilon_\mu
(B^{\dag} \partial^\mu B-\partial^\mu B^{\dag} B),
\end{equation}
to implement the coupling of the $\Upsilon$-field to the B-pseudoscalars.
The coupling constants $g_{\Upsilon B \bar{B}}$ and $f_{\Upsilon}$ are
dimensionless and real.

The naive scaling of the coupling constants $g_{\Upsilon B \bar{B}}$ and
$f_{\Upsilon}$ in the large b-quark mass limit implies \cite{gm}:
\begin{equation}
\label{i20}
g_{\Upsilon B \bar{B}} \to (m_b)^{1/2} \;\;\; {\rm and}\;\;\;
f_{\Upsilon} \to (m_b)^{-1/2}.
\end{equation}
In this limit, the contribution from the $\Upsilon$-states to the r.h.s. in
eq. (13) decouples; and therefore, as was done in ref. \cite{drt}, this
contribution in this limit
can be ignored. However, in the same limit, the residues at
the $\Upsilon$-poles in the B-number form factor in eq. (16) scale as
$(m_b)^2$. If naive scaling holds, then the $\Upsilon$-poles of the b-number
form factor of the B-meson below the
$B \bar{B}$-threshold cannot be neglected, contrary to what was done in the
derivation of the bounds in ref. \cite{drt}. New bounds, if possible, have
to be derived.
\newpage

\noindent
{\bf 3. THE NEW BOUNDS.}
\vspace{0.1in}

The derivation of the new bounds is possible with an appropriate
generalization of the method we already used in \cite{drt}. The
technical details are explained in the appendix. To adapt our problem to
the framework of the appendix we shall map the entire complex $y$-plane
($y={t \over {4M^2_B}}$) onto the unit disc $|z| \leq 1$ via the
transformation
\begin{equation}
\label{i.21}
i {{1+z} \over {1-z}}= \sqrt{y-1}= i \sqrt{ {{1+v.v'} \over 2}}.
\end{equation}
Here $v.v'$ is the Isgur-Wise variable which denotes the product of the
four-velocities of the incoming and outgoing B-mesons in the vertex in
eq.(1): $q^2= 2 M_B^2 (1-v.v')$. Eventually, we are interested in bounds of
the b-number form factor F in the physical region relevant to the decays
in (2), i.e.,
\begin{equation}
\label{i22}
1 \leq v.v' \leq { 1\over 2} (M/M' + M'/M) \simeq 1.6.
\end{equation}
By the transformation in (21), the physical cut $1 \leq y \leq \infty$ is
mapped into the unit circle $z=e^{i \theta}$; the $\bar{B} B$-threshold
at $y=1$ into $z=-1$; and the position of the $\Upsilon$-poles below the
$\bar{B} B$-threshold at $y_i={{M^2_{\Upsilon_i}} \over {4 M^2_B}}$ into
the real points $z_i$:
$-1<z_i<0$; i=1,2,3. The integral in the r.h.s. of eq.(13) can then be
written as an integral on the unit circle.

In order to use the results derived in the appendix, eq.(13) should be cast
into the form
\begin{equation}
\label{i23}
1 \geq {1 \over {2 \pi}} \int^{2 \pi}_0 d \theta |f(e^{i \theta})|^2.
\end{equation}
That this is always possible is guaranteed by the fact that the integrand in
(13) is positive, and the following theorem: let
$\phi (e^{i \theta})$ be real and positive, then
\begin{equation}
\label{i24}
h(z)=exp \left( {1 \over {2 \pi}} \int_0^{2 \pi} d \theta
{{e^{i \theta}+z} \over {e^{i \theta}-z}} \log \phi(e^{i \theta}) \right)
\;,
\end{equation}
verifies
\begin{equation}
\label{i25}
|h(e^{i \theta})|= \phi (e^{i \theta}),
\end{equation}
is analytic and has no zeros in the unit disc. The function $h(z)$ is
unique up to a global phase. This is actually the solution of the
Dirichlet problem of finding an analytic function $h(z)$ with no zeroes
in the unit disc with a boundary condition on the unit circle as given by
(\ref{i25}). The problem can be immediately solved with the help of the
Poisson kernels, applied to the real and imaginary parts of
$\log h(z)$ \cite{mates}.
The solution is given by (\ref{i24}).

It is easy to find directly the function $h$ which corresponds to the
two factors which multiply $|F|^2$ in the integrand in  eq.(13) with the
help of the relations
\begin{equation}
\label{i26}
y-y_i=-4 {{(z-z_i)(1-zz_i)} \over {(1-z_i)^2 (1-z)^2}}=
{{z_i-z} \over {1-z_iz}} \left( {{1+z_i} \over {1-z_i}} +
{{1+z} \over {1-z}} \right)^2,
\end{equation}
where $z$, $z_i$ are the images by (21) of $y$, $y_i$ respectively. Factors
of the type $(z-a)/(1-za^*)$ ($a \in C$) are ubiquitous in the analysis;
they are unimodular on the unit circle and, therefore, drop from the
integrand in eq.(23).

For the sake of simplicity we shall choose $Q^2=0$ in the unitary
inequality in (13) and ignore here the question of optimizing the choice
of $Q^2$. We shall also adopt the lowest order
result in eq. (12) as a good estimate of
the QCD evaluation of $\chi(0)$. Perturbative $\alpha_s$-corrections to this
result are known to be small at the $m_b^2$-scale. Since we shall be ignoring
$\alpha_s$-corrections, which are positive, it seems prudent to neglect
as well the contribution from the $\Upsilon$-states in the r.h.s. of the
unitarity inequality. As already mentioned they decouple in the large
b-quark mass limit in any case.
We shall also take $M_B$ and $m_b$ to be equal. The
unitarity inequality in (12) then reads
\begin{equation}
\label{i27}
1 \geq {{5 n_f} \over {16 N_c}} \int_1^{\infty} dy \; y^{-7/2}
(y-1)^{3/2} |F|^2.
\end{equation}
But for the $n_f$-factor, this coincides with eq.(14) in ref. \cite{drt}.

Equation (27) can now be written in the canonical form of eq.(23), by
setting
\begin{equation}
\label{i28}
f(z)= \varphi(z) F[q^2(z)],
\end{equation}
with
$$
\varphi(z)=\varphi(0) \sqrt{1-z} (1+z)^2 \;\;\; {\rm and}\;\;\;
\varphi (0)= {1 \over {16}} \sqrt{{{5 \pi} \over 2}{{n_f} \over {N_c}}}
\;\;.
$$
\begin{flushright}
(29a,b)
\end{flushright}
As a function of the variable $z$, $F[q^2(z)]$ is an analytic function in
the unit disc, except for the three poles at $-1 < z_i <0$, corresponding
to the locations of the three $\Upsilon$-states below the
$\bar{B} B$-threshold. These poles of $F(q^2)$ give
rise to poles of the function $f(z)$ in (28) at the same location
$-1< z_i <0$, with residues
$$
R_i= \varphi (z_i) {{1-z_i} \over {1+z_i}} z_i \eta_i=
\varphi (0) z_i (1+z_i) (1-z_i)^{3/2} \eta_i \; ; \;\;\; i=1,2,3
$$
\begin{flushright}
(30a)
\end{flushright}
where $\eta_i$ denotes the product of coupling constants
%\newpage
$$ \eta_i \equiv 3 g_{\Upsilon_i B \bar{B}}f_{\Upsilon_i} \;\;.$$
\begin{flushright}
(30b)
\end{flushright}
The modulus of the couplings $|f_{\Upsilon_i}|$ can be determined from
the experimental electronic widths (see eq.(17)). Unfortunately,
the couplings $g_{\Upsilon_i B \bar{B}}$ for the three $\Upsilon$'s below
the $\bar{B} B$-threshold are unknown; and therefore the sizes of the
residues are also unkown. As discussed in the appendix, it is nervertheless
possible to obtain upper and lower bounds on the form factor $F(q^2)$
using the fact that $F(0)=1$ (see eq.(3)), if the locations of the poles
$z_i$ of the function $f(z)$ are known. These are determined, via eq.(21),
by the masses of the three $\Upsilon$-states below the
$\bar{B} B$-threshold; i.e., the parameters
\setcounter{equation}{30}
\begin{equation}
\label{i31}
%z_i={{1-\sqrt{1-a_i}} \over {1+\sqrt{1+a_i}}}; \;\;\;\;
a_i \equiv {{M^2_{\Upsilon_i}} \over {4M^2_B}}.
\end{equation}
We find in this case (see eq.(\ref{a17}) in the appendix)
\begin{equation}
\label{i32}
F_- (z) \leq F(z) \leq F_+ (z),
\end{equation}
where
\begin{equation}
\label{i33}
F_\pm =-{{F(0)} \over {\sqrt{1-z} (1+z)^2}} \prod_i z_i
\left( {{1-z_i z} \over {z-z_i}} \right) \left[
1 \pm \sqrt{{{z^2} \over {1-z^2}}}
\sqrt{ {{512} \over {5 \pi}} {{N_c} \over {n_f}} {1 \over {F^2(0)}}
\prod_i {1 \over {z_i^2}}-1 } \; \right].
\end{equation}
In the case where $a_i \to 1$ ($z_i \to -1$), and correcting for the
$n_f$-factor, the formula coincides with eq.(16) of \cite{drt}. Using the
experimental values for the $\Upsilon$-masses below threshold, we obtain
in particular
upper and lower bounds on the slope of the b-number form-factor of the
B-meson at the origin:
\begin{equation}
\label{i34}
-6.0 \leq F'(v.v'=1) \leq 4.5.
\end{equation}
The lower bound, although rather generous, is not trivial.

As indicated in the appendix, it is also possible
with the same input, to obtain bounds on the $\eta_i$-residues:
\begin{eqnarray}
\label{i35}
& & -3.3 \times 10^3 \leq \eta_1 \leq 3.3 \times 10^3 \nonumber \\
& & -5.7 \times 10^3 \leq \eta_2 \leq 5.7 \times 10^3 \nonumber \\
& & -2.7 \times 10^3 \leq \eta_3 \leq 2.7 \times 10^3 \; .
\end{eqnarray}
As we shall next discuss, these bounds allow for huge values of the unknown
couplings $g_{\Upsilon_i B \bar{B}}$.

In order to get a feeling for what is a reasonable expected size for the
$\eta_i$-residues, we propose to extract the value of the $\eta_4$-residue
corresponding to the $\Upsilon(4S)$ state
which is already above the $4M^2_B$-threshold from experiment.
The experimental data, as well as the corresponding
couplings are shown in Table 1. The decay rate of the $\Upsilon(4S)$ into
$\bar{B} B$ calculated with the effective lagrangian in eq.(19) is
\begin{equation}
\label{i36}
\Gamma (\Upsilon(4S) \to B \bar{B})= {1 \over {48 \pi}}M_{\Upsilon_4}
\left( 1 - {{4 M_B^2} \over {M_{\Upsilon_4}^2}} \right)^{3/2}
g_{\Upsilon_4 B \bar{B}}^2.
\end{equation}
{}From eqs.(\ref{i17}) and (\ref{i36}) and the knowledge of the experimental
total width we obtain for $\eta_4$ in (30b):
\begin{equation}
\label{i37}
|\eta_4 (exp.)| \leq 0.75 \pm 0.15;
\end{equation}
i.e., a value about three orders of magnitude smaller than the limits
allowed by the bounds for the other $\eta_i$, i=1,2,3.

It is also instructive to extract from experiment the corresponding
residues for the charmonium states $\psi(3S)$ and $\psi(4S)$ which are
above the $D \bar{D}$-thres- \break
hold.
The experimental data; as well as the corresponding couplings
are shown in Table 2. It is noteworthy that the experimental upper values
of the $\eta$-residues for the $\psi(3S)$, $\psi(4S)$ states,
and the $\eta_4$-residue of the $\Upsilon(4S)$ state
are of the same size. We propose to use this phenomenological observation
as a guiding ansatz for possible input values of the $\eta_i$-residues
i=1,2,3, and to derive the corresponding bounds for $F(q^2)$
\footnote{We have assumed flavour SU(3) symmetry and used the same
constants $g_{B \bar{B} \Upsilon_i}$, $g_{D \bar{D} \psi_i}$
for any of the three light
flavour species u,d,s of the mesons B and D, respectively.
This allows to improve
the bounds on the residues by including all possible channels. Notice
that not all the channels are always allowed by phase space.}.

The relevant analytic form of the upper and the lower bounds for $F(q^2)$
when both the positions and the modulus of the
residues of the $\Upsilon$-poles are known
is given by eqs.(\ref{a16}) and (\ref{a19}) of the appendix.
In the limit where $|\eta_i| \to 0$ i=1,2,3 we reproduce the first bound
given in \cite{drt} (corrected by the famous $n_f$-factor), i.e.,
\begin{equation}
\label{i38}
-0.89 \leq F'(1)\leq 0.52 \;\;.
\end{equation}
The corresponding upper and lower bounds of the slope $F'(v.v'=1)$ for a
specific set of input values of the modulus of the reduced residues
$|\eta_i|$ of about the same size as the upper bounds
known from experiment are given
in Table 3. We observe from the results in this table that the bounds are
rather insensitive to small variations of the $|\eta_i|$'s. (Increasing
all $|\eta_i|$ by a factor 4, diminishes the lower bound by $50 \%$). The
upper bounds in Table 3 are not interesting since they all have positive
slope and we expect from Bjorken's bound \cite{bj}, that $F(v.v')$ is a
decreasing function for $v.v' \geq 1$. The lower bounds however are
certainly non-trivial and may be useful for phenomenology and model
building.

We conclude from our analysis above that, the only rigorous lower bound we
have at present
on the slope of the b-number form factor of the B-meson at the origin
is the one in eq.(34). Nevertheless, on phenomenological grounds, we
consider that a lower bound
\begin{equation}
\label{i39}
F'(1) \geq -1.7,
\end{equation}
is a conservative estimate.

\newpage
\noindent
{\bf 4. COMMENTS ON THE HEAVY QUARK EFFECTIVE} \break
\noindent {\bf THEORY.}
\vspace{0.1in}

Related to the work described in the previous sections, there are a number
of observations we wish to make, which are relevant to the heavy quark
effective theory formulation \cite{hg} of the Isgur-Wise symmetries.

First we shall comment on the limit $m_b \to \infty$ of our bounds.
As $m_b$ grows bigger and bigger, there appear an increasing number of
$\Upsilon$-resonances below the $\bar{B} B$ threshold (a semiclassical
estimate based on nonrelativistic potential models gives that this
growth goes like $(m_b)^{1/2}$ \cite{qr}).
In order to be able to take the $m_b \to \infty$ limit on our expressions
for the bounds, more information should be known about the $m_b$ dependence
of the location of the poles $z_i(m_b)$, as well as of the residues
$\eta_i(m_b)$.
However, if a behaviour $\eta_i (m_b) \to const.$ and a finite number of
poles below threshold are assumed in the $m_b \to \infty$ limit, as is done in
\cite{dkp}, then, since $z_i(m_b) \to -1$ the results in ref. \cite{drt}
are again recovered. The authors of ref. \cite{dkp} found that the effect
of such poles is to broaden the bounds. This is just an artifact of the
approximation they use. As shown in the appendix, stronger bounds can be
derived leading to the same results as in \cite{drt}.

The second observation is of a phenomenological nature. Based on naive scaling
\cite{gm} of the coupling
constants $f_{\Upsilon}$ and $f_{\psi}$ in the large b-quark mass limit
and the large c-quark mass limit, one expects the ratios of these
couplings to scale as
\begin{equation}
\label{i40}
2 \;{{f_{\Upsilon_i}} \over {f_{\psi_i}}} \to \left( {{m_c} \over {m_b}}
\right)^{1/2},
\end{equation}
where the factor 2 takes care of the different quark charges of the $b$ and
$c$ quarks.
Except for the i=3 states (and in fact the electronic width of the
$\Upsilon(3S)$ is poorly known), the experimental ratios
\begin{equation}
\label{i41}
2\;{{f_{\Upsilon_1}} \over {f_{\psi_1}}} \simeq 0.57; \;
2\;{{f_{\Upsilon_2}} \over {f_{\psi_2}}} \simeq 0.63; \;
2\;{{f_{\Upsilon_4}} \over {f_{\psi_4}}} \simeq 0.69,
\end{equation}
are not incompatible with the empirically allowed range of quark mass
values \cite{pdb}:
\begin{equation}
\label{i42}
0.5 \leq \left( {{m_c} / {m_b}} \right)^{1/2} \leq 0.6.
\end{equation}

Our last comment has to do with the compatibility of the bounds with models
of the Isgur-Wise function. In ref. \cite{drt} we proposed as a simple
minded model of this function the one provided by the triangle graph vertex
with two heavy quark lines and one light quark across with a constituent
mass which acts as a regulator and no gluons across. The resulting
Isgur-Wise function has the form
\begin{equation}
\label{i43}
\xi(v.v'=\omega)={ 1 \over {\sqrt{\omega^2-1}}} \; \log (\omega +
\sqrt{\omega^2 -1}).
\end{equation}
The slope at zero recoil is $\xi'(1)=-1/3$, in confortable agreement with
the lower bound in eq.(\ref{i38}). The function in (\ref{i43})
has been recently found again in a toy field theory model which tries to
implement both heavy and light quark symmetries \cite{bh}. Bardeen and Hill
dismiss however this solution on grounds of
"residual mass invariance" of the
heavy quark effective theory \cite{lm,fnl} and conclude that the Isgur-Wise
function in their model is given by
\begin{equation}
\label{i44}
\xi(v.v'=\omega)={2 \over {1+v.v'}}.
\end{equation}
The slope at zero recoil of this function is $\xi'(1)=-1/2$, also compatible
with the lower bound in (\ref{i38}).

\newpage

\begin{table}
\label{tabbounds}
\caption{Data and coupling constants for the $\Upsilon$ states. The
coupling constants $f_{\Upsilon}$ and $g_{\Upsilon \bar{B} B}$ are
defined by the effective Lagrangians in eqs. (18) and (19).}
\begin{center}
\small
\begin{displaymath}
\begin{array}{|c|c|c|c|c|c|c|}
\hline
State     & Mass              & \Gamma(i \to e^+ e^-)
& |f_{\Upsilon_i}|\times 10^2 &  \Gamma(i \to B \bar{B})
& |g_{\Upsilon_i B \bar{B}}| &|\eta_i|=3 |g_{\Upsilon_i D \bar{D}}
f_{\Upsilon_i}|\\
          & (MeV) & (keV) & & (MeV) & &\\ \hline
\Upsilon(1S) &9460.32 \pm 0.22  &1.34\pm 0.04 &2.5 &  &? &   \\
\Upsilon(2S) &10023.30 \pm 0.31 &0.56         &1.6 &  &? &   \\
\Upsilon(3S) &10355.3 \pm 0.5   &0.44         &1.4 &  &? &   \\
\Upsilon(4S) &10580.1 \pm 3.5   &0.24\pm 0.05  &1.0 &\leq 23.8 \pm 2.2 &\leq 25
&\leq 0.75\\ \hline
\end{array}
\end{displaymath}
\end{center}
\end{table}
%\clearpage
\begin{table}
\label{tabbounds}
\caption{Data and coupling constants for the charmonium states. The
coupling constants $f_{\psi}$ and $g_{\psi \bar{D} D}$ are those
of effective Lagrangians analogous to eqs. (18) and (19).}
\begin{center}
\small
\begin{displaymath}
\begin{array}{|c|c|c|c|c|c|c|}
\hline
State     & Mass     & \Gamma(i \to e^+ e^-)
& |f_{\psi_i}|\times 10^2 &  \Gamma(i \to D \bar{D})
& |g_{\psi_i D \bar{D}}| &|\eta_i|=1.5 |g_{\psi_i D \bar{D}}f_{\psi_i}|\\
          & (MeV) & (keV) & & (MeV) & & \\ \hline
J/\psi(1S) &3096.93 \pm 0.09  &5.35 \pm 0.29 &8.8 &  &? &   \\
\psi(2S) &3686.00\pm 0.10      &2.14 \pm 0.21 &5.1 &  &? &   \\
\psi(3S) &3769.9 \pm 2.5       &0.26 \pm 0.4  &1.8 &\leq 23.6 \pm 2.7
&\leq 16.8 &\leq 0.47\\
\psi(4S) &4040 \pm 10          &0.75 \pm 0.15 &2.9 &\leq 52 \pm 10
&\leq 4.0 &\leq 0.17\\ \hline
\end{array}
\end{displaymath}
\end{center}
\end{table}
\samepage
%\normalsize
%\clearpage
\begin{table}
\label{tablebounds}
Table 3. Upper and lower bounds for the slope of the b-number
form factor of the B-meson for various phenomenological input values of
the residues $\eta_i$ (see eqs. (30b) and (16) in the text).
\begin{center}
\begin{displaymath}
\begin{array}{|c|c|c|c|c|}
\hline
|\eta_1| &|\eta_2| &|\eta_3|  &F'(1)_{lower} &F'(1)_{upper}
\\ \hline
%$0.2$      &$0.2$      &$0.2$       &$-1.03$           &$0.63$ \\
$0.5$      &$0.5$      &$0.5$       &$-1.23$           &$0.79$ \\
$1.0$      &$1.0$      &$1.0$       &$-1.51$           &$1.00$ \\
$1.0$      &$0.5$      &$0.3$       &$-1.36$           &$0.92$ \\
%$0.5$      &$0.3$      &$0.2$       &$-1.15$           &$0.74$ \\
$1.5$      &$1.5$      &$1.5$       &$-1.73$           &$1.04$ \\ \hline
\end{array}
\end{displaymath}
\end{center}
\end{table}
\clearpage

\noindent
{\bf ACKNOWLEDGEMENTS}

\noindent
It is a pleasure to thank G. Anderson, J. Bijnens, D. Broadhurst, H. Dosch,
S. Narison,
M. Neubert, M. Peskin, A. Pich, S. Raby, J. Sloan, J. Soto and
B. Stech for enlightening discussions.
This work was supported in part by USA Grant DE-FG02-91-ER40690.

\clearpage
\renewcommand{\theequation}{A.\arabic{equation}}
\setcounter{equation}{0}
\appendix
\noindent
{\bf  APPENDIX}
\vspace{0.1in}

The mathematical tools needed to derive the bounds on the form factor
$F(q^2)$ in eq.(1) follow from analyticity properties and positivity.

Let $f(z)$ be an analytic function on the unit disc, and let
\begin{equation}
\label{a1}
I[f] \equiv {1 \over {2 \pi}} \int_0^{2 \pi} d \theta
|f( e^{i \theta})|^2=
{1 \over {2 \pi i}} \oint_{|\omega|=1} {{d \omega} \over \omega}
|f(\omega)|^2, \;\; \omega=e^{i \theta}.
\end{equation}
The basic inequality follows from
\begin{equation}
\label{a2}
0 \leq {1 \over {2 \pi}} \int_0^{2 \pi} d \theta
|f( e^{i \theta})-f(0)|^2=
I[f] - |f(0)|^2,
\end{equation}
i.e.,
\begin{equation}
\label{a3}
|f(0)|^2 \leq I[f].
\end{equation}
A similar inequality may be derived at any point $z$ in the unit disc.
With the help of the M\"oebius transformation
\begin{equation}
\label{a4}
\omega={{z-x} \over {1-z^*x}}, \;\;\; |z|^2 <1,
\end{equation}
at fixed $z$, the problem of finding a bound on $|f(z)|$ is reduced to
finding a bound at $x=0$. Indeed, (\ref{a4}) maps the unit circle
$|\omega|=1$ onto the unit circle $|x|=1$, and the point $\omega=z$ is
mapped into $x=0$. In terms of the variable $x$
\begin{equation}
\label{a5}
I[f]=(1-|z|^2) {1 \over {2 \pi i}}\oint_{|x|=1} {{dx} \over x}
\left| { {f( {{z-x} \over {1-z^*x}} )} \over {1-z^*x}} \right|^2,
\end{equation}
for which inequality (\ref{a3}) gives, at $x=0$,
\begin{equation}
\label{a6}
|f(z)|^2 \leq {{I[f]} \over {1-|z|^2}}.
\end{equation}
This is the generalization of (\ref{a3}).

Suppose $f$ has a simple zero in the disc, at $z=a$. One may build
\begin{equation}
\label{a7}
\psi(z) = f(z) {{1-a^* z} \over {z-a}},
\end{equation}
which is analytic on the disc and is such that $|f(z)|=|\psi(z)|$ at
$|z|=1$; therefore $I[\psi]=I[f]$ and applying (\ref{a6}) to $\psi(z)$
yields
\begin{equation}
\label{a8}
|f(z)|^2 \leq {{I[f]} \over {1-|z|^2}} \left|{{z-a} \over {1-a^*z}}\right|^2.
\end{equation}
Notice that
\begin{equation}
\label{a9}
\left|{{z-a} \over {1-a^*z}} \right|^2 = 1 -
{{(1-|a|^2)(1-|z|^2)} \over {|1-a^*z|^2}}<1,
\end{equation}
since $|z|, \; |a| <1$. Knowledge on the location of a simple zero inside
the disc leads to an inequality (\ref{a8}), which is more constraining. The
generalization to higher order zeroes is immediate.
One can proceed similarly when $f(z)$ has a simple pole at $z=p$ in the
unit disc. In that case one may build
\begin{equation}
\label{a10}
\psi(z) = f(z) {{z-p} \over {1-p^* z}},
\end{equation}
which is analytic on the disc. Equation (\ref{a6}) then reads
\begin{equation}
\label{a11}
|f(z)|^2 \leq {{I[f]} \over {1-|z|^2}} \left|{{1-p^* z} \over {z-p}}\right|^2,
\end{equation}
which for a given $I[f]$ is less constrainig than (\ref{a6})
\footnote{Equation (\ref{a11}) is however more constraining than the bounds
obtained in
\cite{dkp}, which would give in this case
$$|f(z)|^2 \leq {{I[f]} \over {1-|z|^2}} \left|{{1-p^* z} \over {z-p}}
\right|^2+
I[f] \left| {{p} \over {z-p}} \right|^2,$$
and for the residue
$$|R|^2 \leq I[f].$$}
. Letting $z \to p$, a bound on the residue $R$ is also obtained
\begin{equation}
\label{a12}
|R|^2 \leq I[f] (1-|p|^2).
\end{equation}

If the residue $R$ is also known, $\psi(z)=f(z)-{R \over {z-p}}$ is
analytic on the disc, eq. (\ref{a6}) applies and yields
\newpage
\begin{equation}
\label{a13}
\left|f(z)-{R \over {z-p}} \right|^2 \leq {{I[\psi]} \over {1-|z|^2}},
\end{equation}
with
\begin{equation}
\label{a14}
I[\psi]=I[f]-{{|R|^2} \over {1-|p|^2}}.
\end{equation}
Notice that in each case the bounds only depend on the information that
is provided about the function.

%One could proceed similarly with other kinds of singularities, e.g.,
%branchcut singularities. Since the strategy is to remove
%the singularity from the function, that would however
%require knowledge on the nature of the branching point.

After these examples, let us consider the cases of interest to us
discussed in the text:

i) $f(0)$ is known.

ii) $f(0)$ is known as well as the location of the poles
$-1 <z_1<z_2<z_3<0$.

iii) The residues $R_1, R_2, R_3$ of these poles are also known.

\vspace{0.1in}
i) Build $\psi(z)=f(z)-f(0)$, for which $\psi(0)=0$. Eq.(\ref{a8}) applies
with $a=0$ and yields
\begin{equation}
\label{a15}
|f(z)-f(0)|^2 \leq {{I[f]-|f(0)|^2} \over {1-|z|^2}} |z|^2.
\end{equation}
This is the inequality used in \cite{drt}.

\vspace{0.1in}
ii) Build
$$\psi(z)={{z-z_1} \over {1-z^*_1 z}}{{z-z_2} \over {1-z^*_2 z}}
{{z-z_3} \over {1-z^*_3 z}}f(z)+z_1 z_2 z_3 f(0);$$
$\psi(0)=0$. The bounds are given by (\ref{a8})
\begin{equation}
\label{a16}
|\psi(z)|^2 \leq {{I[\psi]} \over {1-|z|^2}} |z|^2,
\end{equation}
with
\begin{equation}
\label{a17}
I[\psi]=I[f] - |z_1 z_2 z_3 f(0)|^2.
\end{equation}
Bounds on the residues follow from (\ref{a16}):
\begin{equation}
\label{a18}
\left|{{z_1-z_2} \over {1-z^*_2 z_1}}
{{z_1-z_3} \over {1-z^*_3 z_1}}{{R_1} \over {1-|z_1|^2}}
+z_1 z_2 z_3 f(0)\right|^2 \leq {{I[\psi]} \over {1-|z_1|^2}} |z_1|^2,
\end{equation}
and similarly, {\it mutatis mutandis}, for $R_2$ and $R_3$.

\vspace{0.1in}
iii) Build
$$\psi(z)=f(z)-{{R_1} \over {z-z_1}}-{{R_2} \over {z-z_2}}-
{{R_3} \over {z-z_3}}- \left( {{R_1} \over {z_1}}+{{R_2} \over {z_2}}+
{{R_3} \over {z_3}} + f(0) \right);$$
$\psi(0)=0$. The bounds are given by eq.(\ref{a16}) with
$$I[\psi]=I[f]-\left|{{R_1} \over {z_1}}+{{R_2} \over {z_2}}+
{{R_3} \over {z_3}}+f(0) \right|^2-{{|R_1|^2} \over {1-|z_1|^2}}
-{{|R_2|^2} \over {1-|z_2|^2}}-{{|R_3|^2} \over {1-|z_3|^2}}$$
\begin{equation}
\label{a19}
-2 \; Re \; \left( {{R_1 R_2^*} \over {1-z_1 z_2^*}}+
{{R_1 R_3^*} \over {1-z_1 z_3^*}}+
{{R_2 R_3^*} \over {1-z_2 z_3^*}} \right).
\end{equation}

\end{document}